\documentclass[12pt,aps,nofootinbib]{revtex4}
\usepackage{amsmath}
\usepackage{graphicx}
\usepackage{epsfig}

\renewcommand{\vec}[1]{\boldsymbol #1}

\DeclareMathOperator{\Trace}{Tr}

\DeclareMathOperator{\sign}{sgn}

\begin{document}

\title{Cold asymmetrical fermion superfluids  in   nonperturbative  renormalisation group.}
\author{Boris Krippa}
\affiliation{School  of Physics and Astronomy, University of Manchester, Manchester
M13 9PL, UK}
\date{\today}
\begin{abstract}
The application of the nonperturbative renormalisation group approach to a system 
with two fermion species  is studied. Assuming a simple  ansatz for 
the effective action with effective bosons, describing pairing effects we 
 derive a set of approximate flow equations for 
the effective coupling including boson and fermionic fluctuations. The  case of two fermions with 
different masses but coinciding Fermi surfaces is considered.
 The  phase transition to a 
phase with broken symmetry is found at a critical value of the running 
scale. The large mass difference is found to disfavour the formation of pairs.
  The mean-field results are recovered  if the effects of boson loops are omitted.
While the  boson fluctuation effects  were found to be negligible for large values of $p_{F} a$ they become increasingly important
with decreasing $p_{F} a$ thus making the mean field description less accurate.


\end{abstract}
\maketitle
 The properties of asymmetric many fermion systems have recently attracted much attention (see, for example Ref. \cite{He} and references therein)
 driven by the substantial advance in experimental studies of trapped fermionic atoms. This asymmetry can be provided by unequal masses, different densities and/or
chemical potentials. Understanding   the pairing mechanism in such settings  would  be of immense value for different many fermion systems
from atomic physics to strongly interacting quark matter. The important theoretical issue  to be resolved here  is the nature of the ground state. Several 
competing states have been proposed so far. These include: LOFF \cite{Lar} phase, breached-pair (BP)
superfluidity \cite{Wil} (or Sarma phase) and mixed phase \cite{Bed}. Establishing  the true ground state  is still an open question.  
It was shown, for example,   that LOFF and mixed  phases are more stable then the Sarma phase in the systems of fermions with the mismatched Fermi surfaces 
and with both equal and different masses \cite{He, Bed, Cal}. All these studies, however, have been performed within the mean field approximation (MFA).
In spite of the fact that in many cases MFA is quite reliable it is important to  understand better the limits of applicability of MFA
and work out the physical regimes where  the MFA is too crude or even inadequate.  The convenient way to estimate the corrections to MFA is
provided by the nonperturbative renormalisation group (NRG) approach \cite{Wet} which was successfully applied to 
the standard pairing problem with one type of fermions \cite{Kri, Kr1, Kr2, Kr3}. The main element of NRG is the effective average action $\Gamma_k$
which is a generalisation of the standard effective action $\Gamma$, the generating functional of the 1PI Green functions. The only difference between 
them is that $\Gamma_k$ includes only quantum fluctuations with momenta larger then the infrared scale $k$. The evolution of the system
as the function of the scale $k$ is described by the nonperturbative flow equations. When $k\rightarrow 0$ all fluctuations
are included and full effective action is recovered. Similarly, at starting scale $k=K$ no fluctuations are included so 
$\Gamma_{k=K}$ can be associated with the classical action $S$ therefore  $\Gamma_k$ provides an interpolation between the classical and 
full quantum effective actions. 

The dependence of $\Gamma_k$ from the infrared scale $k$ is given by the nonperturbative renormalisation group equation (NRGE)

\begin{equation}
\partial_k\Gamma=-\frac{i}{2}\,\Trace \left[(\partial_kR)\,
(\Gamma^{(2)}-R)^{-1}\right].
\end{equation}
Here $\Gamma^{(2)}$ is the second functional derivative of the effective action  taken with respect to all types of field included in the action
and $R(q,k)$ is a regulator which should suppress the contributions of states with momenta less than or of the order of running scale $k$.
To recover the full effective action we require $R(q,k)$ to vanish as $k\rightarrow$ 0 whereas for $q << k$ the regulator behaves as
$R(q,k) \simeq k^2$. The above written equation is, in general, the functional equation. For a practical applications it needs to be  converted  to the 
system of partial or ordinary differential equations so that approximations and truncations are required. 

We consider a nonrelativistic  many-body system at zero temperature  with two types of  the fermion species $a$ and $b$ interacting 
through a short-range attractive interaction and 
introduce a boson field $\phi$ describing the pair of interacting fermions. The ansatz for $\Gamma$ takes the form

\begin{eqnarray}
\Gamma[\psi,\psi^\dagger,\phi,\phi^\dagger,\mu,k]&=&\int d^4x\,
\left[\phi^\dagger(x)\left(Z_\phi\, (i \partial_t +\mu_a + \mu_b)
+\frac{Z_m}{2m}\,\nabla^2\right)\phi(x)-U(\phi,\phi^\dagger)\nonumber\right.\\
\noalign{\vskip 5pt}
&&\qquad\qquad+{\sum_{i=a}^{b}}\psi_{i}^\dagger\left( Z_{\psi,i} (i \partial_t+\mu_i)
+\frac{Z_{M,i}}{2M_i}\,\nabla^2\right)\psi_i\nonumber\\
\noalign{\vskip 5pt}
&&\qquad\qquad\left.- Z_{g} \left(\frac{i}{2}\,\psi_{b}^{\rm T}\sigma_2\psi_{a}\phi^\dagger
-\frac{i}{2}\,\psi_{a}^\dagger\sigma_2\psi_{b}^{\dagger{\rm T}}\phi\right)\right].
\label{eq:ansatz}
\end{eqnarray}
Here $M_i$ is the mass of the fermion in vacuum and the factor $1/2m$ with $m=M_a + M_b$ in the 
boson kinetic term is chosen simply to make $Z_m$ dimensionless. The
coupling $Z_g$,  the 
wave-function renormalisations factors $Z_{\phi,\psi}$ and the kinetic-mass 
renormalisations factors $Z_{m,M}$ all run with on $k$, the scale of the regulator. Having in mind the future applications to the 
 crossover from BCS to BEC (where chemical potential becomes negative) we also let the chemical potentials $\mu_a$ and $\mu_b$ run,
 thus keeping the corresponding densities (and Fermi momenta $p_{F,i}$) constant.  
The bosons are , in principle, coupled to the chemical potentials via a quadratic term in $\phi$, 
but this can be absorbed into the potential by defining $\bar U=U-(\mu_1 +\mu_2) Z_\phi\phi^\dagger\phi$. 
We expand this potential about its minimum, $\phi^\dagger\phi=\rho_0$, so that the coefficients $u_i$ are defined at $\rho=\rho_0$,  
\begin{equation}
\bar U(\rho)= u_0+ u_1(\rho-\rho_0)
+\frac{1}{2}\, u_2(\rho-\rho_0)^2
+\frac{1}{6}\, u_3(\rho-\rho_0)^3+\cdots,
\label{eq:potexp}
\end{equation}
where we have introduced $\rho=\phi^\dagger\phi$. Similar expansion can be written for the renormalisation factors.
 The coefficients of the expansion run with the scale.  The phase of the system is determined by the
 coefficient $u_1$.
We start evolution at high scale where the system is in the symmetric phase so that  $u_1 > 0$. When the running scale  becomes 
comparable with the pairing scale (close to  average Fermi-momentum) the system undergoes the phase transition to the phase with broken symmetry,
energy gap etc. The point of the transition corresponds to the scale where   $u_1 = 0$. The bosonic excitations in the gapped phase are 
gap-less Goldstone bosons. Note, that  in this phase the minimum of the potential will also run with the scale $k$ so that the value  $\rho_0 (k \rightarrow 0)$
determines the physical gap.

The evolution equation takes the following  general form

\begin{equation}
\partial_k\Gamma=-\frac{i}{2}\,\Trace \left[(\partial_kR_B)\,
(\Gamma^{(2)}_{BB}-R_B)^{-1}\right] + \frac{i}{2}\,\Trace \left[(\partial_kR_F)\,
(\Gamma^{(2)}_{FF}-R_F)^{-1}\right].
\end{equation}
Here $\Gamma^{(2)}_{BB(FF)}$ is the matrix of the second functional derivatives of the effective action  taken with respect to boson(fermion)  fields included 
in the action
and $R_B (R_F)$ is the boson (fermion) regulator which should suppress the contributions of states with momenta
less than or of the order of running scale $k$.  The boson regulator  has the structure 
\begin{equation}
  R_B = R_B diag(1,1). 
\end{equation}
The fermion regulator for both types of fermions 
has the structure 
\begin{equation}
 R_{F,i} = sgn(\epsilon_{i}(q) - \mu_i) R_{F,i}(q,\mu_i,k) diag(1,-1)
\end{equation}
 Note that this regulator is positive for particle states above the Fermi surface and negative for the hole states below the Fermi surface.

The evolution equations include running of chemical potentials, effective potential and all couplings ($Z_\phi, Z_{m}, Z_{M,i}, Z_{\psi,i}, Z_g$).
 However, in this paper we allow to run only $Z_\phi$, parameters in the effective potential ($u's$ and $\rho_0$) and 
chemical potentials since this is the minimal set needed to include the effective boson dynamics.

Calculating the second functional derivatives, taking the matrix trace  and carrying out the pole integration in the loop integrals
 we get the evolution equation for $U$ at constant chemical potentials
\begin{eqnarray}
\partial_k \bar U=-\frac{1}{{\cal V}_4}\,\partial_k\Gamma
&=&-\frac{1}{2}\,\int\frac{d^3{\vec q}}{(2\pi)^3}\,\frac{E_{F,S}}
{\sqrt{E_{F,S}^2+\Delta^2}}\,[\sign(q - p_{\mu,a})\,\partial_kR_{F,a} + \sign(q - p_{\mu,b})\,\partial_kR_{F,b}]\nonumber\\
\noalign{\vskip 5pt}
&&+\,\frac{1}{2Z_\phi}\int\frac{d^3{\vec q}}{(2\pi)^3}\,
\frac{E_{BR}}{\sqrt{E_{BR}^2-V_B^2}}
\,\partial_kR_B.\label{eq:potevol}
\end{eqnarray}
Here 
\begin{equation}
E_S = (E_{F,a} +E_{F,b})/2,\qquad E_A = (E_{F,a} - E_{b,a})/2,
\end{equation}
 and 
\begin{equation}
E_{B}(q,k)=\frac{Z_m}{2m}\,q^2+u_1
+u_2(2\phi^\dagger\phi-\rho_0)+R_B(q,k), \qquad V_B= u_2\phi^\dagger\phi,
\end{equation} 
\begin{equation}
E_{F,i}(q,p_{\mu,i},k)=\frac{1}{2 M_i}\,q^2-\mu_i+R_F(q,k)\,\sign(q-p_{\mu,i}),  \qquad \Delta^2=g^2\phi^\dagger\phi.
\label{eq:defEFR}
\end{equation}
and we have introduced
$p_{\mu,i}=\sqrt{2M_{i}\mu_i}$, the Fermi momentum corresponding to the (running) value of $\mu_i$.
It is worth mentioning that poles 
 in the fermion  propagator occur at
\begin{equation}
q_{0}^{1,2}= - E_A \pm\sqrt{E_{S}(q,k)^2+\Delta^2}.
\end{equation}
At $k=0$ ($R_F=0$) in the condensed phase, these become
 exactly the dispersion relations obtained in \cite{Wil} where the possibility of having the gapless excitations
has been discussed. The ordinary BCS spectrum can easily be recovered when the asymmetry of the system is vanishing
($E_{A}\rightarrow 0$). 
The first term in the evolution equation for the effective potential describes the evolution of the system related to the  fermionic degrees of
 freedom whereas the second one takes into
 account the bosonic contribution. The mean field results can be recovered if the second term
 is omitted. In this case the equation
for the effective potential can be integrated analytically. 
\begin{equation}
\bar U(\rho,\mu,k)=\bar U(\rho,\mu,K)-\int\frac{d^3{\vec q}}{(2\pi)^3}
\left[{\sqrt{E_{S}(q,k)^2+\Delta^2}}
-{\sqrt{E_{S}(q,K)^2+\Delta^2}}\right]\,.
\end{equation}
At  starting scale $K$ the potential  has the form
\begin{equation}
\bar U(\rho,\mu,K)=u_0(K)+u_1(K)\,\rho.
\end{equation}
The renormalised value of $u_1(K)$ can be related to the scattering
length.
\begin{equation}
\frac{u_1(p_F,K)}{g^2}=-\,\frac{M}{2\pi a}
+\frac{1}{2}\,\int\frac{d^3{\vec q}}{(2\pi)^3}\,
\left[\frac{1}{E_{S}(q,0,0,0)}-\frac{1}{E_{S}(q,\mu_a,\mu_b,K)}\right]\, .
\label{eq:u1Kfull}
\end{equation}
Here $M$ is the reduced mass and the dependence of $E_{S}$ on the chemical potentials has been made explicit.

Differentiating the effective potential with respect to $\rho$, setting the derivative to zero and taking the limit
$K\rightarrow \infty$, we arrive at the equation 
\begin{equation}
-\,\frac{M}{2\pi a}+\frac{1}{2}\int\frac{d^3{\vec q}}{(2\pi)^3}\,
\left[\frac{1}{E_{S}(q,0,0,0)}-\frac{1}{\sqrt{E_{S}(q,\mu_a,\mu_b,k)^2+\Delta^2}}
\right]=0.
\end{equation}
 Taking the physical limit ($k=0$) we obtain the gap 
equation identical to that derived in the mean field approximation \cite{Bed, He}.

We now turn to the full set of the evolution equations which includes the effects of the bosonic fluctuations. In this paper we consider the case of two 
fermion species with the different masses and the  same Fermi momenta.
It implies that the chemical potentials are different. In this situation the Sarma phase does not exist and the system experiences the BCS pairing depending
however on the mass asymmetry. The general case of the mismatched Fermi surfaces will be discussed in the subsequent publication.

The derivation of the evolution equations was discussed in details in Ref. \cite{Kri} so that here we just mention the main points.
Within the above described approximation  (fixed couplings $Z_{m}, Z_{M,i}, Z_{\psi,i}, Z_g$) all of these can be obtained from the evolution of 
the effective potential, for example
\begin{equation}
Z_\phi = -\,\frac{1}{2}\left.\frac{\partial^2}{\partial \mu\partial\rho}
\Bigl(\partial_k \bar U\Bigr)\right|_{\rho=\rho_0},
\end{equation}
where $\mu = \mu_a + \mu_b$. Substituting the expansion for the effective potential on the left-hand side of the evolution equation leads to a set of 
ordinary differential equations for the running minimum $\rho_0$ and coefficients $u_n$. These equations have a generic form  
\begin{equation}
\partial_k u_n -u_{n+1}\partial_k\rho = \,\left.\frac{\partial^n}{\partial\rho^n}
\Bigl(\partial_k \bar U\Bigr)\right|_{\rho=\rho_0},
\label{eq:Zphialt}
\end{equation}
One can see from this equation that some sort of closure approximation is needed as the equation for $u_n$ always include $u_{n+1}$ coefficient etc.
In this paper we calculated $u_{n>2}$ in the MFA with the   effective potential given by the Eq.(12).
 As already mentioned we follow the evolution of the chemical potential keeping density fixed. Defining the total derivative 
\begin{equation}
\frac{d}{dk}=\partial_k+\frac{d\rho_0}{dk}\,\frac{\partial}{\partial\rho_0}.
\end{equation}
and applying it to the $\frac{\partial\bar U}{\partial\rho}$ (or to $\frac{\partial\bar U}{\partial\mu})$ we obtain the following set of equations
\begin{equation}
-2z_{\phi 0}\,\frac{d\rho_0}{dk}+\chi\,\frac{d\mu}{dk}
=-\left.\frac{\partial}{\partial \mu}
\Bigl(\partial_k \bar U\Bigr)\right|_{\rho=\rho_0},
\label{eq:muevol}
\end{equation}
where $z_{\phi 0}$ is the coefficient in the leading term of the expansion for $Z_\phi$ similar to Eq.(3),
and
\begin{eqnarray}
\frac{du_0}{dk}+n\,\frac{d\mu}{dk}
&=&\left.\partial_k \bar U\right|_{\rho=\rho_0},\\
\noalign{\vskip 5pt}
-u_2\,\frac{d\rho_0}{dk}+2z_{\phi 0}\,\frac{d\mu}{dk}
&=&\left.\frac{\partial}{\partial \rho}
\Bigl(\partial_k \bar U\Bigr)\right|_{\rho=\rho_0},\\
\noalign{\vskip 5pt}
\frac{du_2}{dk}-u_3\,\frac{d\rho_0}{dk}+2z_{\phi 1}\,\frac{d\mu}{dk}
&=&\left.\frac{\partial^2}{\partial \rho^2}
\Bigl(\partial_k \bar U\Bigr)\right|_{\rho=\rho_0},\\
\noalign{\vskip 5pt}
\frac{dz_{\phi 0}}{dk}-z_{\phi 1}\,\frac{d\rho_0}{dk}+\frac{1}{2}\,\chi'\,
\frac{d\mu}{dk}
&=&-\,\frac{1}{2}\left.\frac{\partial^2}{\partial \mu\partial\rho}
\Bigl(\partial_k \bar U\Bigr)\right|_{\rho=\rho_0},
\end{eqnarray}
where we have defined
\begin{equation}
\chi'=\left.\frac{\partial^3\bar U}{\partial \mu^2\partial\rho}
\right|_{\rho=\rho_0}, \qquad z_{\phi 1}=-\left.\frac{1}{2}
\frac{\partial^3 U}{\partial\mu\partial^{2}\rho}\right|_{\rho=\rho_0}.
\end{equation}
These functions have also been calculated in the MFA.
The set of evolution equations in symmetric phase can easily be recovered 
using the fact that chemical potential does not run in symmetric phase and that
$\rho_0 =0$.

Let us now turn to the results. For simplicity we consider the case of the hypothetical ``nuclear'' matter with short range attractive interaction 
between  two types of fermions, light and heavy,
and study the behaviour of the energy gap as the function of the mass asymmetry. We choose the Fermi momentum to be  $p_{F}=1.37 fm^{-1}$.
 One notes that the formalism is applicable to any type of a many-body system with two fermion species
 from quark matter to fermionic atoms so that the    hypothetical asymmetrical ``nuclear'' matter is simply chosen as  a study case.
We assume that $M_a < M_b$, where $M_a$ is always the mass of the physical nucleon.

 In this paper we use a sharp cutoff function  chosen in the form 
 which makes the loop integration  as simple as possible
\begin{equation}
R_{F,i}=\frac{k^2}{2M_i}\left[((k + p_{\mu,i})^{2}- q^2)\theta(p_{\mu,i}+k -q) + ((k + p_{\mu,i})^{2}+ q^2 -2p^{2}_{\mu,i})\theta(q - p_{\mu,i}+k)\right],
\end{equation}
and similarly for the boson regulator
\begin{equation}
R_{B}=\frac{k^2}{2m}(k^2 - q^2)\theta(k -q). 
\end{equation}
Here $\theta(x)$ is the standard step-function.
This type of  boson regulator was also used in Ref. \cite{Bla} (see also Ref.\cite{Litim}).  

The use of a sharp cutoffs can be potentially dangerous as it may generate the artificial singularities when calculating the flow of the  
renormalisation constants $(Z's)$ but seem to be harmless when all the evolution parameters are related to the effective potential RG flow as is the case here.

As we can see the fermion sharp cutoff consists of two terms which result in modification of the particle  and hole propagators
 respectively. 
The hole term is further modified to suppress the contribution from the surface terms, which may bring in the dangerous dependence
 of the regulator on the cutoff scale even at the vanishingly small $k$. We found that the value of the gap practically does not depend
on the starting point provided $M_{a,b} << K$. As expected, the system undergoes the phase transition
to the gapped phase at some critical scale which depends on the value assumed for the parameter $p_{F} a$ where $a$ is the scattering length in vacuum.
 One notes that the critical scale does not depend on the mass asymmetry.

First we consider the case of the unitary limit where the scattering length $a = -\infty$. 
The results of our calculations for the gap 
 are shown on Fig. 1. 

\begin{figure}
\begin{center}
\includegraphics[width=8cm,  keepaspectratio,clip]{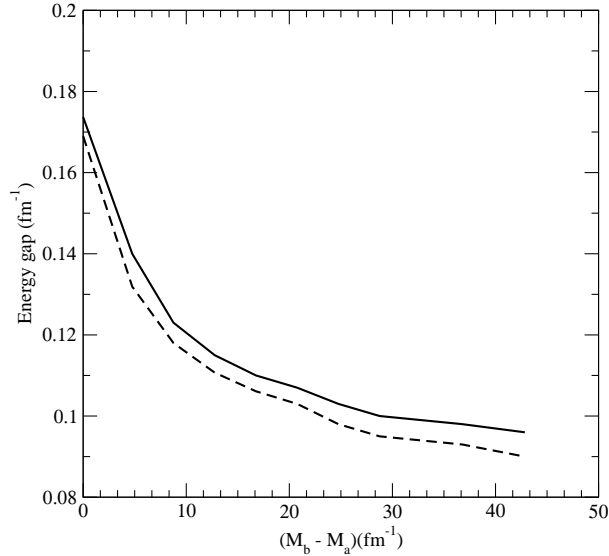}
\end{center}
\caption{\label{fig:running1}Evolution of the gap in the MF approach (dashed curve) and with boson loops (solid curve) in the unitary regime
$a = -\infty$ as a function of a mass asymmetry.}
\end{figure}

 We see from this figure  that  increasing  mass asymmetry leads to a decreasing gap that seems to be a natural result.
However, the effect of the boson loops is found to be   small. We found essentially no effect in symmetric phase, $2-4 \%$ corrections for the value of the 
gap in the broken phase and even smaller corrections for the chemical potential so that one can conclude that
the MF approach indeed provides the reliable description in the unitary limit for both small and large mass asymmetries.
It is worth mentioning that the
 boson contributions are more important for the evolution of $u_2$ where they drive  $u_2$ to zero as $k\rightarrow 0$ making the effective potential
convex in agreement with the general expectations. This tendency  retains in the unitary regime regardless of the mass asymmetry. 

We have also considered
the behaviour of the gap as the function of the parameter $p_F a$ for the cases of the zero asymmetry $M_a = M_b$ and the  maximal asymmetry $M_b = 10 M_a$.
The results are shown on Fig.2.
\begin{figure}
\begin{center}
\includegraphics[width=8cm,  keepaspectratio,clip]{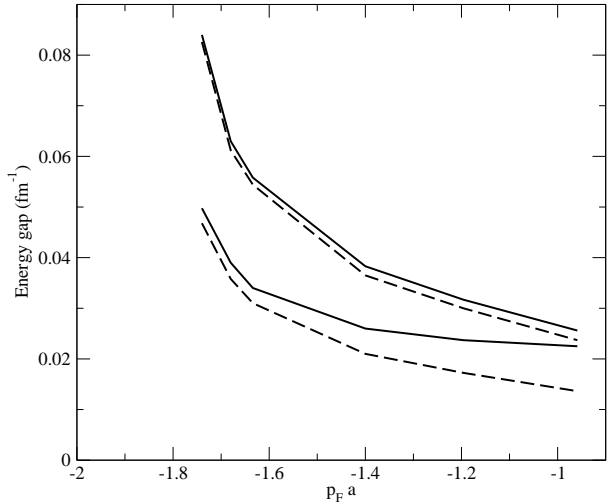}
\end{center}
\caption{\label{fig:running2}Evolution of the gap as a function of the parameter $p_F a$. The upper 
pair of the  curves corresponds to the calculations with no asymmetry in the MF approach (dashed curve) and with boson loops (solid curve) and the lower pair 
of the curves describes the results of calculations with the  maximal asymmetry when $M_b = 10 M_a$}.
\end{figure}

  One can see from  Fig.2 that in the  case of zero (or small) asymmetry  the corrections stemming from  boson loops
are small at all values of   the parameter $p_F a$ considered here (down to $p_F a = 0.94)$. On the contrary, when  $M_b = 10 M_a$ these corrections, being 
rather small at $p_F a \geq  2$ becomes significant ($\sim  30\%$) when the value of $p_F a$ decreases down to $p_F a \sim  1$. We found that
at $p_F a \sim  1$ the effect of boson fluctuations becomes non negligible, $\sim  10\%$  already for $M_b = 5 M_a$. One can therefore conclude 
 that the regime of large mass asymmetries, which starts  approximately at  $M_b > 5 M_a$, 
 moderate scattering length and/or the Fermi momenta is the one where the MF description becomes less accurate so that the calculations going beyond the MFA
are needed. One might expect that the deviation from the mean field results could even be stronger in a general case of a large mass asymmetry and 
the mismatched Fermi surfaces but the detailed conclusion can only be drawn after the actual calculations are performed.

 We were not able to follow the evolution of the system at
 small gap (or small $p_F a$) because of the non-analyticity of the effective action
 in this case which means that the power expansion of the effective potential around the minimum is
no longer reliable. To find the evolution at small gap the partial differential equation for the effective potential should probably be solved. 

To summarise, we have studied the pairing effect for the asymmetric fermion matter with two fermion species as a function of fermion mass asymmetry.
 We found that  regardless of the size of the fermion mass asymmetry
the boson loop corrections are small at large enough values  of $p_F a$ so that  the MFA provides a consistent description of the pairing effect 
in this case. However, when $p_F a \sim  1$ these corrections become significant at large asymmetries ($M_b > 5 M_a$) making the MFA inadequate. 
In this case it seems to be necessary to go beyond the mean field description.

There are several ways where this approach can further be developed. The next natural step  would be to consider the case of the mismatched Fermi surfaces
 taking into account the possibility of formation of Sarma, mixed and/or LOFF phases  and exploring the importance of the boson loop for the stability of those phases
and applying the approach to the real physical systems like fermionic atoms, for example.
Work in this direction is in progress. The other important extension of this approach would be to include running of all
couplings of the effective action and use different type of cut-off function, preferably the smooth one. The three body force effects \cite {Kr4},
 when the correlated pair interact
with the unpaired fermion may also turn out important, especially for non-dilute systems.

The author  thanks Mike Birse, Niels Walet and Judith McGovern for numerous very helpful discussions.

\end{document}